# LLM-guided phase diagram construction through high-throughput experimentation

Ryo Tamura[1,2,3*], Haruhiko Morito[4*], Yuna Oikawa[2], Guillaume Deffrennes[5], Shoichi Matsuda[6], Naruki Yoshikawa[1], Tomoaki Takayama[7,8], Taichi Abe[9], Koji Tsuda[2,1,3], and Kei Terayama[10,11*]

*1 Center for Basic Research on Materials, National Institute for Materials Science, 1-1 Namiki, Tsukuba, Ibaraki 305-0044, Japan*

*2 Graduate School of Frontier Sciences, The University of Tokyo, 5-1-5 Kashiwa-no-ha, Kashiwa, Chiba 277-8561, Japan*

*3 RIKEN Center for Advanced Intelligence Project, 1-4-1 Nihonbashi, Chuo-ku, Tokyo 103-0027, Japan*

*4 Core Facility Center, Tohoku University, 2-1-1 Katahira, Aoba-ku, Sendai 980-8577, Japan*

*5 Univ. Grenoble Alpes, CNRS, Grenoble INP, SIMaP, F-38000 Grenoble, France*

*6 Center for Green Research on Energy and Environmental Materials, National Institute for Material Science, 1-1 Namiki, Tsukuba, Ibaraki 305-0044, Japan*

*7 Graduate School of Science and Technology, Nara Institute of Science and Technology, Ikoma, Nara 630-0192, Japan*

*8 Data Science Center, Nara Institute of Science and Technology, Ikoma, Nara 630-0192, Japan*

*9 Research Center for Structural Materials, National Institute for Materials Science, 1-2-1 Sengen, Tsukuba, Ibaraki 305-0047, Japan*

*10 Graduate School of Medical Life Science, Yokohama City University, 1-7-29, Suehiro-cho, Tsurumi-ku, Kanagawa 230-0045, Japan*

*11 MDX Research Center for Element Strategy, Tokyo Institute of Technology, 4259 Nagatsuta-cho, Midori-ku, Yokohama, Kanagawa, 226-8501, Japan*

**Abstract**

Constructing phase diagrams for multicomponent alloys requires extensive experimental measurements and is a time-consuming task. Here we investigate whether large language models (LLMs) can guide experimental planning for phase diagram construction. In our framework, a general-purpose LLM serves as the experimental planner, suggesting compositions for measurement at each cycle in a closed loop with high-throughput synthesis and X-ray diffraction phase identification. Using this framework, we experimentally constructed the ternary phase diagram of the Co–Al–Ge system at 900 °C through iterative synthesis and characterization. We compared two strategies that differ in how the initial compositions are selected: one uses predictions from a domain-specific LLM trained on phase diagram data (aLLoyM), while the other relies solely on the general-purpose LLM. The two strategies exhibited complementary strengths. aLLoyM directed the initial measurements toward compositionally complex regions in the interior of the ternary diagram, enabling the earliest discovery of all three novel phases that form only in the ternary system. In contrast, the general-purpose LLM adopted a textbook-like approach which efficiently identified a larger number of phases in fewer cycles. In addition, a simulated benchmark comparing the LLM against conventional machine learning confirmed that the LLM achieves more efficient exploration. The results demonstrate that LLMs have high potential as experimental planners for phase diagram construction.

## Introduction

The rapid advance of artificial intelligence (AI) is transforming the landscape of scientific research, a paradigm now widely referred to as "AI for Science." Within this paradigm, large language models (LLMs) have emerged as particularly versatile tools for accelerating discovery across the physical sciences[1–4]. General-purpose LLMs such as GPT-4, Gemini, and Claude possess broad scientific knowledge and have been utilized for a wide range of tasks, including experimental design, scientific reasoning, and code development. Recent studies have explored LLMs as autonomous agents for experimental planning, where LLMs integrate literature knowledge, analyze data, and propose informed next experiments, which is expected to accelerate materials developments[5–9].

One research area where efficient experimental planning is important is the construction of phase diagrams. Although phase diagrams are fundamental to materials design, their experimental determination remains costly and time-consuming, particularly for ternary and higher-order systems where the compositional space grows combinatorially. A complete ternary isothermal section may require hundreds of measurements, and as a result, many phase diagrams for ternary and higher-order systems remain unreported. To address this challenge, machine-learning-driven experimental planning methods have been developed to reduce the number of experiments required for phase diagram construction[10–13]. For phase diagrams comprising multiple discrete phase regions, an active learning method based on uncertainty sampling known as the Phase Diagram Construction (PDC) algorithm was developed[14–17], and a web application version, AIPHAD, is publicly available[18]. However, these machine learning models require measured data to train on, necessitating substantial initial measurements before the surrogate model becomes informative. Moreover, conventional machine learning models require phase names to be converted into numerical labels, which introduces an additional preprocessing step.

LLMs, with their broad scientific knowledge, may offer a complementary approach not only by providing informed initial guidance without requiring training data, but also by serving as autonomous experimental planners that integrate multiple information sources throughout the exploration process. In contrast to conventional machine learning models, LLMs can handle phase names directly as text without converting them into numerical labels, which simplifies the workflow for phase diagram construction. Although LLMs have been explored for phase diagram-related tasks such as phase prediction[19,20] and diagram annotation[21], their use as autonomous experimental planners that iteratively guide composition selection in a closed loop with experiments remains largely unexplored.

In this work, we investigate how LLMs can guide experimental planning for phase diagram construction. We applied our LLM-based framework to the Co–Al–Ge ternary system, whose phase diagram has not been previously reported and whose ternary compounds are absent from the large-scale DFT database of the Materials Project[22]. The constructed phase diagram therefore represents both a testbed for evaluating LLM-guided exploration and a new contribution to the phase diagram literature. We entrusted experimental planning to a general-purpose LLM (Claude, Anthropic), which suggested compositions for measurement at each cycle. According to these suggestions, high-throughput synthesis was performed and phases were identified using X-ray diffraction (XRD) measurements (Fig. 1). This closed loop between LLM-driven composition selection, synthesis, and characterization was iterated over six cycles. We compared two strategies that differ in how the initial compositions are selected: one uses predictions from aLLoyM, a domain-specific LLM for phase diagrams[20], to guide the first cycle, while the other relies solely on the general-purpose LLM throughout. aLLoyM is a Mistral-Nemo-based model fine-tuned on the Computational Phase Diagram Database (CPDDB)[23], capable of predicting equilibrium phases for arbitrary alloy compositions and temperatures. Although aLLoyM's predictions are approximate and use a standardized nomenclature based on structural prototypes (e.g., B2, C36, and FCC_A1) and generic labels (e.g., SOLID and LIQUID) rather than system-specific phase names, they provide prior information about likely phase distributions. Our high-throughput experiments driven by LLMs revealed that the choice of initial compositions has a decisive impact on the subsequent exploration trajectory, and the two strategies, with and without aLLoyM, exhibited complementary strengths. When guided by aLLoyM, the initial measurements targeted compositionally complex regions in the interior of the ternary diagram, leading to the immediate discovery of novel phases that form only in the ternary system. In contrast, the general-purpose LLM adopted a textbook-like approach, that is, first anchoring the corners and binary edges before moving inward, which efficiently identified a larger number of phases in fewer cycles, which resulted in faster identification of a larger number of phases. These results indicate that the domain-specific LLM is advantageous for discovering novel phases, while the general-purpose LLM excels at efficiently mapping the overall phase distribution. In addition, a simulated benchmark comparing the LLM against PDC uncertainty sampling and random sampling confirmed that the LLM achieved more effective exploration overall, further supporting the potential of LLMs as experimental planners for phase diagram construction. Based on these results, we implemented and released the LLM-EP (LLM-based Experimental Planner) module in the open-source NIMO package, enabling researchers to readily apply LLM-guided experimental planning to their own systems.

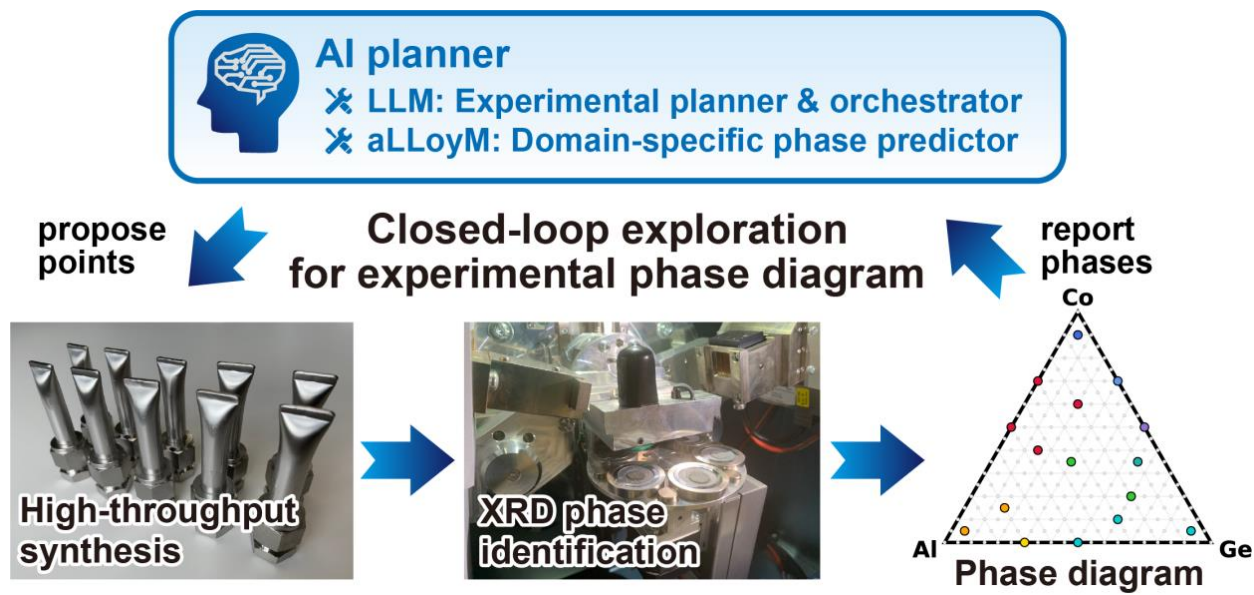

**Fig. 1** Closed-loop workflow for experimental phase diagram construction orchestrated by LLMs. At each cycle, the AI planner, consisting of a general-purpose LLM and a domain-specific LLM (aLLoyM), selects compositions for measurement. High-throughput synthesis and X-ray diffraction (XRD) phase identification are then performed, and the experimental results are fed back to the AI planner for the next cycle.

# Results

**Experimental design**

We investigated the Co–Al–Ge ternary system at 900 °C (1173.15 K), defining 231 candidate compositions on a 5 at.% grid across the ternary section. Each experimental planning strategy was allocated six cycles of eight measurements each (48 total), with composition selection performed by a general-purpose LLM through Claude Code (Anthropic)[24], an agentic coding tool that autonomously reasons about the experimental data and, as needed, executes analysis code through multi-step interactions. In each cycle, a majority-vote protocol was employed: ten independent selection runs per cycle, with the eight most frequently selected compositions chosen for measurement. Notably, in each independent run, the LLM spontaneously adopted distinct exploration philosophies without explicit instructions to do so. For example, some runs prioritized phase diversity by targeting the centers of unexplored regions, others focused on refining suspected phase boundaries by placing points between two known distinct phase regions, and still others adopted a space-filling approach to maximize spatial coverage. The majority-vote aggregation then distilled these diverse perspectives into a balanced set of compositions suitable for effective batch experiments. We compared two strategies that differ in how the initial compositions are selected:

- Strategy A: In the first cycle, a general-purpose LLM selects compositions with reference to aLLoyM phase predictions. From the second cycle onward, the LLM selects compositions based on its own scientific knowledge and all previously obtained experimental results.
- Strategy B: A general-purpose LLM selects compositions based solely on its internal materials science knowledge and previously obtained experimental results throughout all six cycles.

These Strategies were conducted in separate LLM sessions with separate working directories to prevent information cross-contamination. The aLLoyM prediction results are summarized in Fig. S1.

**Exploration results of Co–Al–Ge experimental phase diagram**

The cycle-by-cycle evolution of the experimental phase diagram constructed by each strategy through high-throughput synthesis and XRD measurements is summarized in Fig. 2. In these phase diagrams, a total of 11 phases were confirmed. When coexisting phases were observed in a single sample, only the primary phase was used to label that composition, as detailed in Supplementary Note A. Apparently, Strategy A explores the interior of the ternary section, while Strategy B samples more diverse regions. To investigate the features of each strategy, the cycle dependence of quantitative metrics is examined. The number of discovered phases as a function of cycle is summarized in Fig. 3(a). Strategy B discovered phases more rapidly in the early cycles, reaching 9 phases by cycle 2 and all 11 phases by cycle 4. Strategy A was slower in this regard, reaching 11 phases at cycle 6. This difference arises because the general-purpose LLM in Strategy B adopted a space-filling approach from the first cycle, efficiently covering diverse regions of the composition space, while Strategy A's aLLoyM-guided selection concentrated measurements in the interior of the ternary diagram. From explorations using both strategies, the Co–Al–Ge system at 900 °C contains three phases which form only in the ternary system not present in existing phase diagram databases: B20 Co(Al/Ge) and $Co_2(Al/Ge)_3$, which were reported in Ref.[25], and an unidentified phase designated X, whose details are summarized in Supplementary Note B. The discovery timing of these novel phases reveals a striking contrast between the two strategies (Fig.

3(b)). Strategy A discovered B20 Co(Al/Ge) in the initial selection (cycle 1), whereas Strategy B required cycle 3. The X phase was found by Strategy A in cycle 2, compared to cycle 4 for Strategy B. $Co_2(Al/Ge)_3$ was identified by both strategies in cycle 1. Critically, Strategy A achieved complete discovery of all three novel phases within just 16 measurements (cycle 2), whereas Strategy B required 32 measurements (cycle 4). This result demonstrates that aLLoyM's guidance toward compositionally complex regions in the interior of the ternary diagram was decisive for the early discovery of novel phases which form in the ternary system. At the same time, it is noteworthy that Strategy B, relying solely on the general-purpose LLM without any domain-specific tool, achieved faster overall phase coverage, highlighting the strong potential of LLMs as autonomous experimental planners for phase diagram construction.

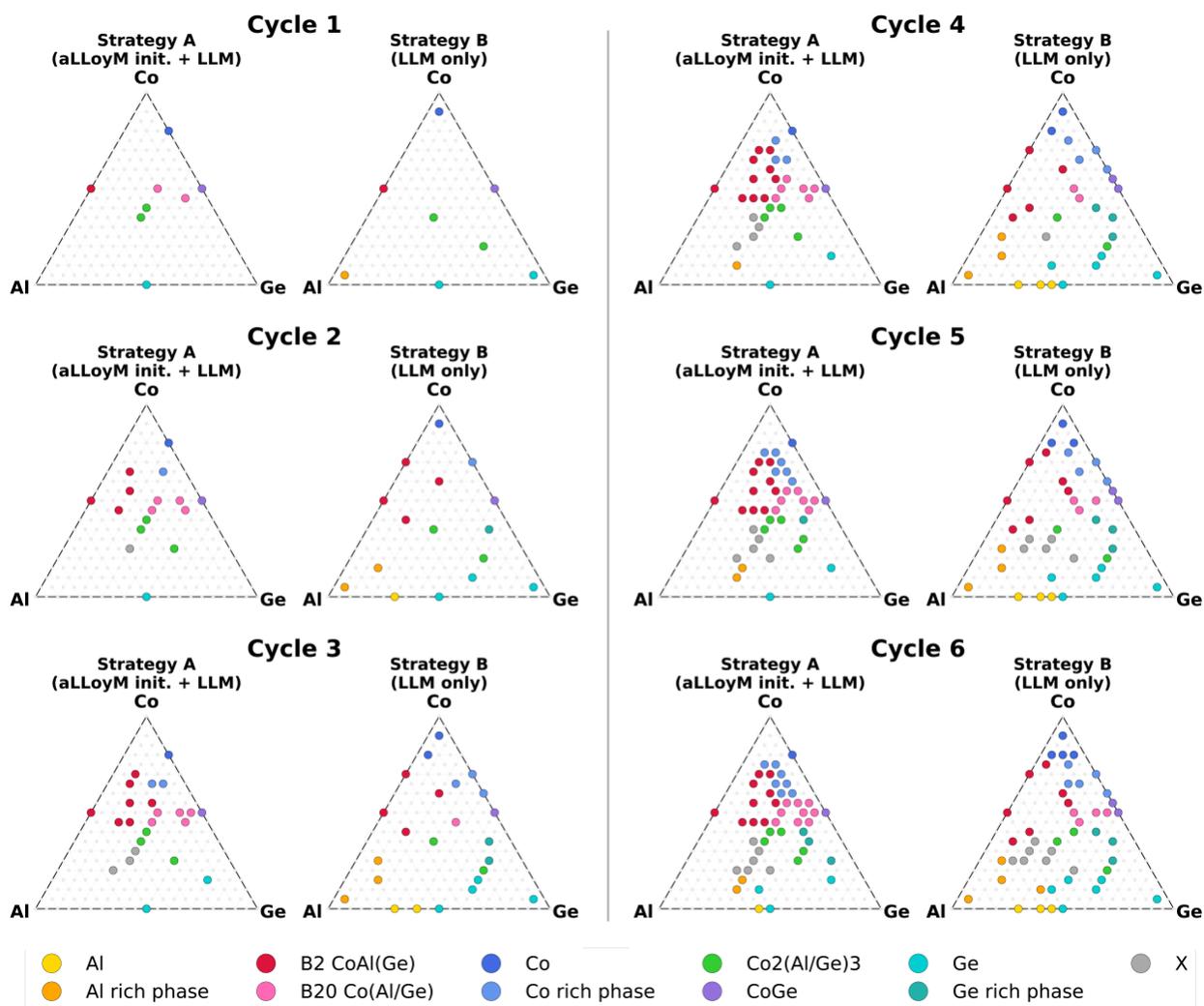

**Fig. 2** Cycle-by-cycle evolution of the experimental phase diagrams constructed through high-throughput synthesis and XRD phase identification, guided by Strategies A (left) and B (right). Strategy A preferentially explores the interior regions of the ternary diagram, while Strategy B covers diverse regions starting from the corners and edges.

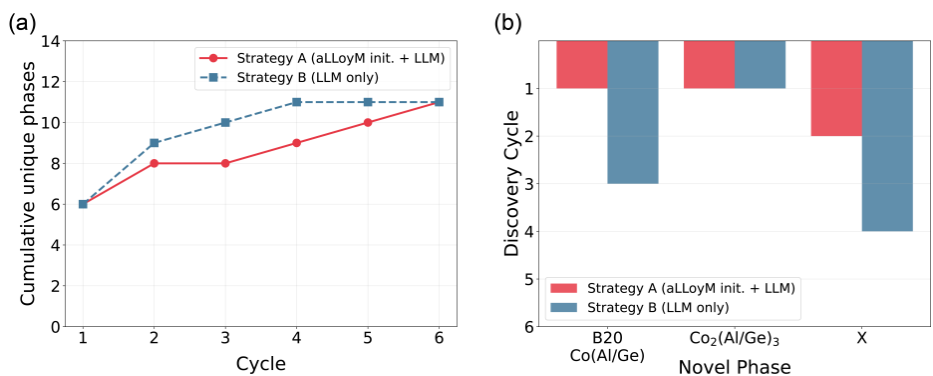

**Fig. 3** Experimental exploration results for the Co–Al–Ge system at 900 °C obtained through high-throughput synthesis and XRD phase identification. (a) Cumulative number of unique phases discovered as a function of cycle. (b) Cycle at which each novel phase was first discovered. Lower values indicate earlier discovery.

**Differences in initial selection with and without aLLoyM guidance**

We summarize the initial selections for each strategy to deepen our understanding of the LLM's behavior as an experimental planner. The initial compositions selected by the LLM with reference to aLLoyM predictions (Strategy A), along with the LLM's selection rationales, are summarized in Table 1. The aLLoyM-guided initial selection led to the discovery of novel phases at four out of eight points. To understand this outcome, we examine the LLM's selection rationales summarized in Table 1. At Co: Al: Ge = 40: 30: 30, aLLoyM predicted the Laves C36 phase, and the measurement revealed $Co_2(Al/Ge)_3$, a novel ternary intermetallic compound. Similarly, at Co: Al: Ge = 50: 20: 30, the prediction of C30 Laves coexisting with FCC_A1 led to the discovery of B20 Co(Al/Ge). At Co: Al: Ge = 35: 35: 30, aLLoyM predicted the unusual CQG_B2 phase notation, which appears only once in the entire 231-point prediction set, and this point also yielded $Co_2(Al/Ge)_3$. Although none of these CALPHAD phase names correspond to the actual phases discovered, all four points where novel phases were found had been flagged by aLLoyM as hosting complex multi-phase equilibria. It is also notable that at Co: Al: Ge = 45: 10: 45, where aLLoyM predicted simply "SOLID" with no further phase specification, the LLM selected this point for its high information gain due to the ambiguity of the prediction, and B20 Co(Al/Ge) was discovered.

The spatial distribution of the initial selections reveals a clear contrast between the two strategies. The aLLoyM-guided selection did not target any of the three corners of the ternary diagram, instead placing four of eight points in the interior where multiple elements are mixed in comparable proportions. The remaining four points were placed on binary edges. In contrast, the general-purpose LLM in Strategy B placed three points near the corners (Co-rich, Al-rich, and Ge-rich), three on binary edge midpoints, and only two in the interior. The four compositions common to both strategies, namely Co: Al: Ge = 50: 50: 0, 50: 0: 50, 0: 50: 50, and 35: 35: 30, are all either binary edge midpoints or near the centroid. The aLLoyM-guided strategy replaced the corner-anchoring points with interior measurements targeting phases predicted to form only in the ternary system, which, as shown in the previous section, resulted in the earlier discovery of novel phases. As shown in Table 1, the LLM's selection rationales for Strategy A indicate that it appropriately prioritized compositions where aLLoyM predicted complex or ambiguous phase equilibria. Similarly, the initial selections in Strategy B closely resemble what an experimentalist would choose as a starting point for an unknown ternary system.

**Table 1** Initial compositions selected by the LLM in Strategy A, with reference to aLLoyM predictions. The table lists each composition, the phases predicted by aLLoyM, the LLM's rationale for selecting that point, and the primary phase observed in the experiment.

| Composition Co: Al: Ge | aLLoyM Prediction | Selection Rationale by LLM | Observed Primary Phase (experiments) |
|---|---|---|---|

| 80: 0: 20 | SOLID + FCC_A1 + HCP_A3 | Unique HCP prediction; only point in dataset; validates FCC/HCP boundary on Co-Ge edge | Co |
| --- | --- | --- | --- |
| 40: 30: 30 | SOLID + C36 | Laves C36 phase center; near centroid of triangle; important intermetallic | $Co_2(Al/Ge)_3$ |
| 50: 20: 30 | SOLID + C30 + FCC_A1 | Rare C30 Laves + FCC; boundary between Laves and FCC regions | B20 Co(Al/Ge) |
| 0: 50: 50 | SOLID + LIQUID | Al-Ge binary edge midpoint; anchors the low-Co region | Ge |
| 50: 50: 0 | SOLID + B2 | Co-Al binary edge midpoint; validates B2 ordered phase | B2 CoAl |
| 45 : 10 : 45 | SOLID | Maximally ambiguous prediction; unknown phase — high information gain | B20 Co(Al/Ge) |
| 50 : 0 : 50 | SOLID | Co-Ge binary edge; ambiguous prediction — high information gain | CoGe |
| 35 : 35 : 30 | SOLID + CQG_B2 + SOLID | Unique CQG_B2 notation (only 1 point); near centroid; validates unusual prediction | $Co_2(Al/Ge)_3$ |

**Obtained experimental phase diagram**

The experimental phase diagram of the Co-Al-Ge system has not been reported. Here, unique observed points were collected from two strategies, and the resulting phase diagram at 900 °C is shown in Fig. 4(a). In total, 78 over 231 compositions on a 5 at.% grid across the ternary section were identified by experiments. In addition, the full phase diagram predicted by the PDC algorithm implemented in the NIMO package[26] using the label propagation method is shown in Fig. 4(b). Three novel phases spread broadly around the center of the ternary section, where the three constituent elements are mixed in comparable proportions. It should be noted that this diagram represents the distribution of phases obtained experimentally and does not correspond to an equilibrium phase diagram at 900 °C. In an equilibrium phase diagram, only the thermodynamically stable phases at 900 °C should be shown. Given that aluminum and germanium have melting points of 660 °C and 850 °C, respectively, the presence of a liquid phase would also be expected at this temperature. In contrast, the present diagram reflects the primary phase obtained after heating mixtures of Co, Al, and Ge powders with specified nominal compositions at 900 °C, followed by cooling to room temperature. Therefore, it represents the primary phase assemblages formed under these specific experimental conditions rather than equilibrium states.

The newly discovered X phase occupies a relatively large region in the Al-rich area. Based on single-crystal XRD analysis, the X phase was identified as a Co-Al-Ge ternary cubic phase with space group $Im\bar{3}m$. The details of crystal structure information of the X phase are shown in Supplementary Note B.

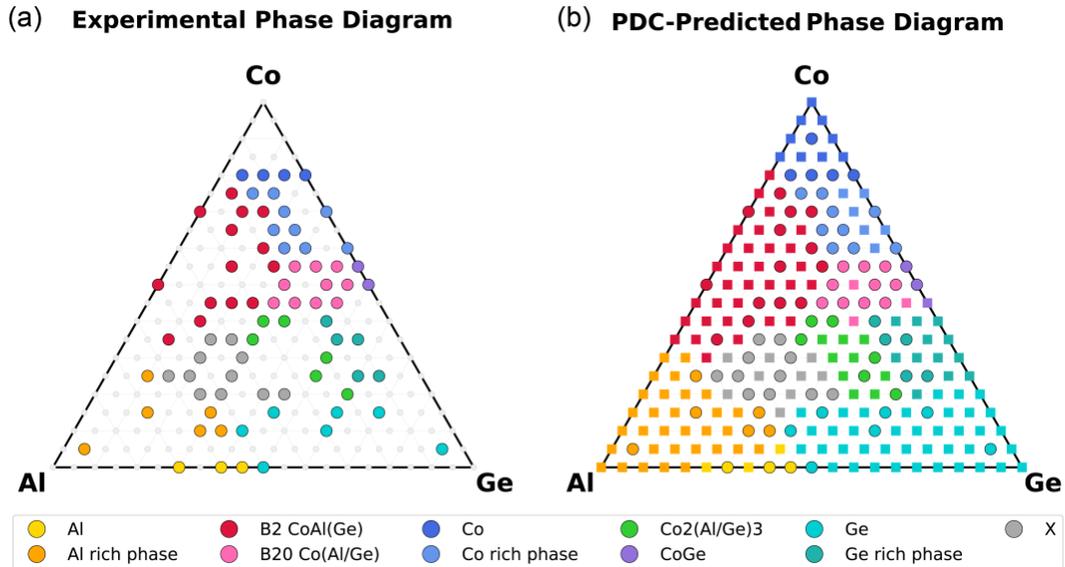

**Fig. 4** Phase diagrams of the Co-Al-Ge system at 900 °C. (a) Experimental phase diagram constructed by combining all measured compositions from Strategies A and B (78 out of 231 compositions). (b) Complete phase diagram predicted by the PDC algorithm using label propagation based on the experimental data. Predictions were made using the phase with the highest probability. Circles indicate experimentally measured compositions, and squares indicate PDC-predicted compositions.

**Comparison with other sampling methods**

To validate the effectiveness of LLM-guided composition selection for phase diagram construction, we compared it against other sampling methods. Since the complete phase diagram of the Co–Al–Ge system was obtained through label propagation applied to our experimental data (Fig. 4(b)), this predicted phase diagram can serve as a ground-truth dataset for benchmarking different experimental planning methods using simulated experiments, where proposed compositions are immediately assigned their known phases without actual synthesis.

For this benchmark, we implemented a new LLMEP (LLM-based Experimental Planner) module in the autonomous experimentation platform NIMO package[27] that enables automated composition proposals via the Claude API[28]. This allows systematic comparison of three methods under identical conditions: LLM-based selection, PDC uncertainty sampling implemented in the NIMO package, and random sampling. Five independent trials of six cycles (eight compositions per cycle) were conducted. In each trial, the LLM independently selected its own initial compositions based on its scientific knowledge, while PDC and random sampling shared the same randomly selected initial eight compositions, since these methods have no prior knowledge to inform the initial selection. We evaluated performance using three metrics: the number of discovered phases as a function of cycle, the discovery timing of the three novel phases, and the Macro F1 score computed by applying label propagation to the sampled data and comparing the predicted full phase diagram against the ground truth.

An important consideration in this benchmark is the difference between interactive LLM usage and programmatic API access. In Strategies A and B for experimental exploration (Fig. 2), Claude Code was employed as an autonomous agent capable of iterative reasoning, code execution, and multi-step decision-making. In contrast, the LLMEP module in NIMO accesses the Claude API through a single prompt-response interaction. Preliminary tests confirmed that API-based proposals differ substantially from those generated through interactive Claude Code sessions, particularly when integrating external information from aLLoyM. We therefore adopted the Strategy B protocol (without aLLoyM) for the LLM benchmark in the comparison, as the general-purpose LLM's scientific reasoning based on available experimental results is more robustly reproduced through API calls than the complex multi-tool workflow of Strategy A. The difference between agentic and API-based LLM usage represents an important consideration for future implementations.

The results are summarized in Fig. 5. The LLM-based selection outperformed both PDC and random sampling in the number of discovered phases and the Macro F1 score throughout the exploration process. Regarding the discovery timing of novel phases, the LLM achieved the fastest detection of $Co_2(Al/Ge)_3$ and the X phase. However, for B20 Co(Al/Ge), PDC discovered this phase earlier than the LLM. The PDC performance metrics in Fig. 5(a) follow a similar improvement trend to the LLM but with a consistent offset. The rate of phase diagram refinement by LLM and PDC is comparable, but PDC starts behind the LLM due to a less informative initial selection. These results demonstrate the strong potential of LLM-guided experimental planning for phase diagram construction, as the LLM consistently outperformed both PDC uncertainty sampling, which is specifically designed for this task, and random sampling in overall exploration efficiency.

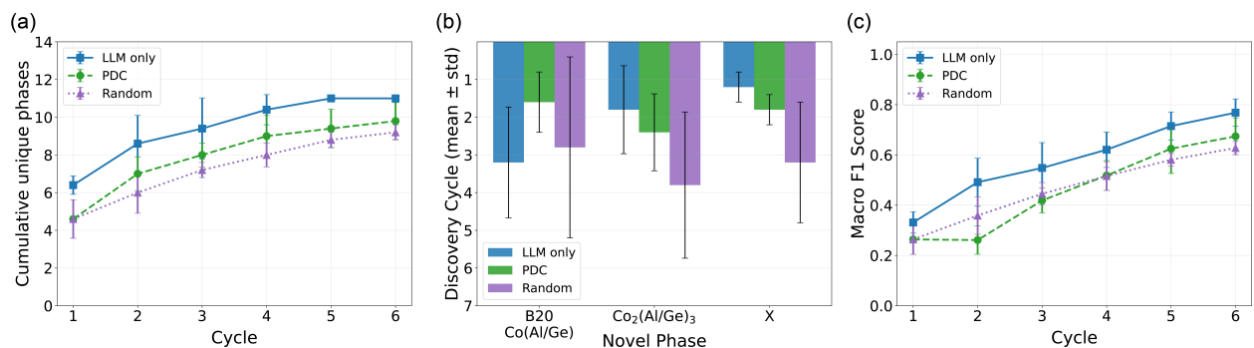

**Fig. 5** Comparison of sampling methods using simulated experiments based on the predicted phase diagram (Fig. 4(b)) as ground truth. Five independent trials of six cycles were conducted for each method: LLM-based selection (Strategy B), PDC uncertainty sampling, and random sampling. (a) Cumulative number of unique phases discovered as a function of cycle. (b) Cycle at which each novel phase was first discovered. (c) Macro F1 score of the full phase diagram predicted by label propagation from the sampled data, compared against the ground truth. Error bars indicate the standard deviation over five trials.

## Discussion

In this work, we demonstrated that a general-purpose LLM can serve as an autonomous experimental planner for phase diagram construction. In the Co–Al–Ge ternary system at 900 °C, we experimentally identified 11 phases, including three novel phases, through six cycles of LLM-guided high-throughput synthesis and XRD measurements. The two strategies, with and without aLLoyM, exhibited complementary strengths. aLLoyM enabled the earliest discovery of novel phases, while the general-purpose LLM achieved faster identification of a larger number of phases through a textbook-like approach. Furthermore, a simulated benchmark confirmed that LLM-guided selection outperformed both PDC uncertainty sampling and random sampling in overall exploration efficiency. These results collectively demonstrate that LLMs possess strong capability as autonomous experimental planners for phase diagram construction.

Interestingly, aLLoyM's predictions use a standardized nomenclature based on structural prototypes and generic labels, rather than system-specific phase names such as B20 Co(Al/Ge) or $Co_2(Al/Ge)_3$. Nevertheless, aLLoyM identified compositional regions with complex multi-phase equilibria, predicting coexistence of Laves phases (C36 and C30) and unusual ordered phases (CQG_B2) in the central portion of the ternary diagram. This indirect guidance directed the initial measurements toward precisely those regions where novel phases reside, even though the specific phases were not predicted by name. This finding suggests that domain-specific LLMs need not provide exact predictions to be valuable; approximate knowledge of phase complexity across composition space is sufficient to guide efficient exploration.

An important direction for future work is the extension to higher-order systems (quaternary and beyond) and the construction of temperature-dependent phase diagrams. These tasks require a substantially larger number of experiments, making the integration with self-driving laboratories[29–31], where synthesis, characterization, and AI-driven planning operate in a fully automated closed loop, increasingly important for realizing scalable, autonomous phase diagram construction. In addition, combining LLMs with other machine learning methods for phase diagram construction, such as the PDC algorithm and other domain-specific LLMs, is a promising direction. In this work, the general-purpose LLM served as the sole decision-maker after the initial cycle, but further efficiency gains may be achieved by allowing the LLM to dynamically orchestrate multiple tools throughout the exploration process. For example, in higher-order systems where the candidate space grows rapidly, PDC could first narrow down the candidate compositions based on uncertainty, and the LLM could then select the most informative measurements from this reduced set, combining the strengths of uncertainty-driven and knowledge-driven approaches.

## Methods

**High-throughput experiments**

The samples for identifying the products with high-throughput experiments were synthesized by direct solid-liquid reaction of Co, Al, and Ge powders[32,33]. Co powders (Alfa Aesar Co., Ltd., 99.8% purity; mean particle size 1.6 μm), Al powders (Kojundo Chemical Laboratory Co., Ltd., 99.99%, < 45 μm), and Ge powders (Kojundo Chemical Laboratory Co., Ltd., 99.99%, < 45 μm) were weighed in the prescribed molar ratios and subsequently mixed in an agate mortar. The mixed powder was pelletized into a columnar shape (ø 5 mm diameter, ~1 mm height) using a die and placed in a boron nitride (BN) crucible (Zikusu Industry Co., Ltd., 99.7%, 8.5 mm outer diameter, 6.5 mm inner diameter, 18 mm depth) in air. Then, in an Ar-filled glove box ($O_2$ and $H_2O$ < 1 ppm), the crucible containing the sample was placed in a one-end-welded stainless-steel tube (SUS316, 12.7 mm outer diameter, 10.5 mm inner diameter, 80 mm height) and sealed with a stainless-steel cap (SUS316) (see Fig. 1). A photograph and schematic diagram of this stainless-steel container have been previously reported[34]. To synthesize the samples, this container was heated at 900 °C for 24 h in an electric furnace in an air atmosphere. The crystalline phases present in the synthesized samples were identified through powder XRD, utilizing a Bruker D2-Phaser system with CuKα radiation at 30 kV and 10 mA. Examples of phase identification by XRD are presented in Supplementary Note A.

**aLLoyM prediction**

Phase predictions for all 231 compositions at 900 °C were obtained using aLLoyM, a Mistral-Nemo-Instruct (12B parameter) model fine-tuned with LoRA adapters on the CPDDB. Although a Model Context Protocol (MCP) server for aLLoyM is available on Hugging Face

(https://huggingface.co/spaces/Playingyoyo/aLLoyM), in this study we pre-generated the phase diagram of the Co–Al–Ge system using aLLoyM and utilized it directly. Predictions were generated on Google Colab using an NVIDIA T4 GPU with 4-bit quantization via the unsloth framework (https://github.com/NIMS-DA/aLLoyM_pred).

**LLM-based experimental planning for real experiment**

Claude Opus 4.6 (claude-opus-4-6) via Claude Code (Anthropic, 2026) was used as an AI agent for composition selection. A key advantage of using an LLM for this task is that phase names can be provided and interpreted directly as text, without requiring conversion into numerical labels. In each cycle, Claude Code autonomously read input files, analyzed the data, and executed multi-step reasoning to select compositions. The agent was prompted with: (i) the list of 231 candidate compositions, (ii) experimentally measured phases for previously selected compositions, and (iii) aLLoyM predictions which were summarized as jsonl format. The agent performed 10 independent selection runs of 8 compositions each, and the final 8 were determined by majority vote across all runs. The prompts are summarized in the Supplementary Note C.

**LLMEP module implemented in NIMO**

To enable systematic use of LLMs as experimental planners for phase diagram construction, we implemented a new LLMEP (LLM-based Experimental Planner) module in the NIMO package[27]. This module interfaces with the Claude API (Anthropic) to generate composition proposals based on its own scientific knowledge and available experimental results. The module can be invoked as follows:

```
nimo.selection(method = "LLMEP",
               input_file = "candidates.csv",
               output_file = "proposals.csv",
               num_objectives = 1,
               num_proposals = 8,
               prompt_file = "prompt.md",
               system_prompt_file = "system.md",
               model = "claude-opus-4-6",
               num_runs = 10,
               log_file = "lm_log.md",
               api_key = "api_key")
```

The candidate compositions and any previously measured phases are stored in candidates.csv, and the proposed compositions are written to proposals.csv. Two prompt files are required. The system_prompt_file defines the LLM's role and general instructions, such as acting as an expert materials scientist specializing in phase diagram construction. The prompt_file provides the task-specific context, including the target system, candidate compositions, and experimental data from previous cycles. The actual prompts used in this work are provided in Supplementary Note D. The LLM model can be selected from available Claude models via the model parameter. The num_runs parameter specifies the number of independent selection runs for the majority-vote protocol, where each run generates num_proposals compositions and the most frequently selected compositions across all runs are chosen as the final proposals. The complete LLM interaction log is saved to log_file for reproducibility and analysis. Using the experimental history provided in candidates.csv and the two prompt files, the module automatically constructs a single API prompt and retrieves the LLM's composition proposals. The Claude API key is specified via the api_key parameter.

It should be noted that the LLMEP module is a general-purpose tool, not limited to phase diagram construction. By providing appropriate prompts via the prompt_file and system_prompt_file parameters, the module can be applied to a wide range of experimental planning tasks. Furthermore, this API-based module operates through a single prompt-response interaction, in contrast to the agentic Claude Code sessions used in the main experiments (Strategies A and B), where the LLM could iteratively reason, execute code, and refine its decisions. This distinction is discussed further in the Results section.

**PDC algorithm setting**

For uncertainty-based phase diagram construction, the PDC algorithm implemented in the NIMO package was used. Label propagation was employed as the classification model to predict phase labels for unmeasured compositions based on the spatial distribution of measured data. The margin sampling strategy was used to compute the uncertainty score for each unmeasured candidate, defined as the difference between the two highest predicted class probabilities. Lower margin values indicate higher uncertainty, meaning the model is less confident in its phase assignment. Batch selection of multiple compositions per cycle was performed by ranking candidates according to their uncertainty scores and selecting the top-ranked unmeasured compositions.

## Acknowledgement

We would like to strongly acknowledge Kaori Yoshida and Yoko Tachibana for experimental contributions. Prof. Hisanori Yamane supported the crystal structure analysis. This study was supported by a project subsidized by JSPS KAKENHI (25K01492 and 25KJ0870), MEXT Program: Data Creation and Utilization Type Material Research and Development Project (JPMXP1122715503, JPMXP1122712807, and JPMXP1122683430), and JST PRESTO (JPMJPR24T8).

## Corresponding author

Correspondence should be addressed to Ryo Tamura (tamura.ryo@nims.go.jp), Haruhiko Morito (haruhiko.morito.b5@tohoku.ac.jp), Kei Terayama (terayama@yokohama-cu.ac.jp).

## Author contributions

R.T., H.M., and Ke.T. conceived the idea and designed the research. R.T. performed the numerical experiments, and H.M. performed the experiments. R.T., H.M., T.T., and Ke.T. discussed the use of LLMs for phase diagram construction. R.T., Y.O., G.D., T.A., and Ko.T. developed aLLoyM and discussed its use in the LLM-based selection. R.T., S.M., N.Y., and Ko.T. developed NIMO and discussed the implementation of the LLM module in NIMO. R.T. wrote the draft, and all authors revised the manuscript.

## Competing interests

All authors declare no financial or non-financial competing interests.

## Data availability

All data generated during this study are included in this published article and its supplementary information files.

## Code availability

NIMO is available at https://github.com/NIMS-DA/nimo. aLLoyM model weights are available at https://huggingface.co/Playingyoyo/aLLoyM.

**Supplementary Figure**

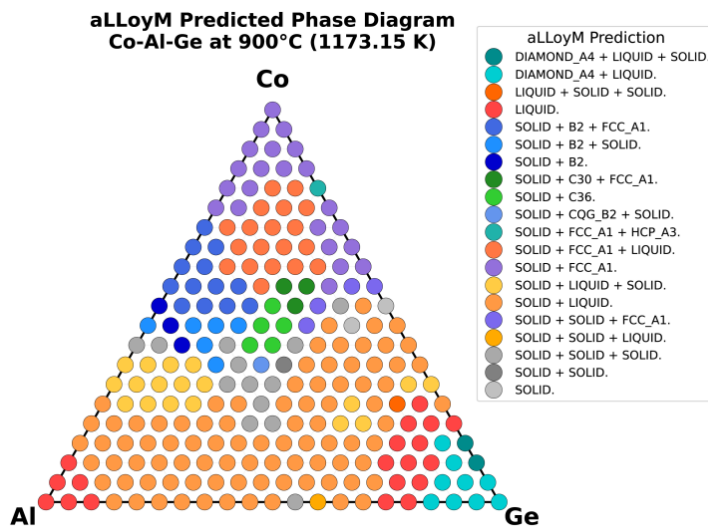

**Fig. S1** Ternary phase diagram of the Co-Al-Ge system at 900 °C predicted by aLLoyM. The 231 candidate compositions on a 5 at.% grid are colored according to the predicted phases. aLLoyM outputs a standardized nomenclature based on crystal structure types and generic labels (e.g., B2, FCC_A1, SOLID, and LIQUID), which differs from system-specific phase names used in the experimental identification.

**Supplementary Note A: Phase identification of formed phases by X-ray diffraction (XRD)**

In this study, phase identification of the formed phases was carried out by X-ray diffraction (XRD) measurements using powder samples. Figure S2 presents representative examples of phase identification for ternary synthesis experiments proposed by the LLM. For the sample synthesized at Co: Al: Ge = 40: 30: 30, it was confirmed that a single phase, which can be indexed as $Co_2(Al/Ge)_3$, was formed. For Co: Al: Ge = 50: 20: 30, the B20 phase was identified; however, diffraction peaks corresponding to a secondary phase, B2-type CoAl(Ge), were also observed in the XRD pattern. In Table 1, B20 Co(Al/Ge) is listed as the "Observed Primary Phase," as the B20 phase was successfully obtained. Similarly, for Co: Al: Ge = 45: 10: 45 and 35: 35: 30, the B20 phase and $Co_2(Al/Ge)_3$ were identified as the primary phases, respectively. As described above, primary and secondary phases coexist in each sample. In the phase diagram shown in Fig. 4, only the primary phases are presented.

When a sample was synthesized with a composition of Co: Al: Ge = 20: 40: 40, peaks that could not be indexed were observed; these are referred to as the X phase in this paper. The powder XRD pattern of the sample obtained under this condition is shown in Fig. S3. As shown in Fig. S3 (a), unreacted Ge remains in the sample, making it appear at first glance that only Ge is present. However, as shown in Fig. S3 (b), magnification of the XRD pattern reveals several small peaks. Characteristic peaks of the X phase are observed around $2\theta = 19°$, $30°$, and $45°$. Samples exhibiting these peaks are identified as the X phase and are indicated as such in the phase diagram. In addition, since single crystals were found in this sample, crystal structure analysis of the X phase was carried out using those single crystals. The details are described in Supplementary Note B.

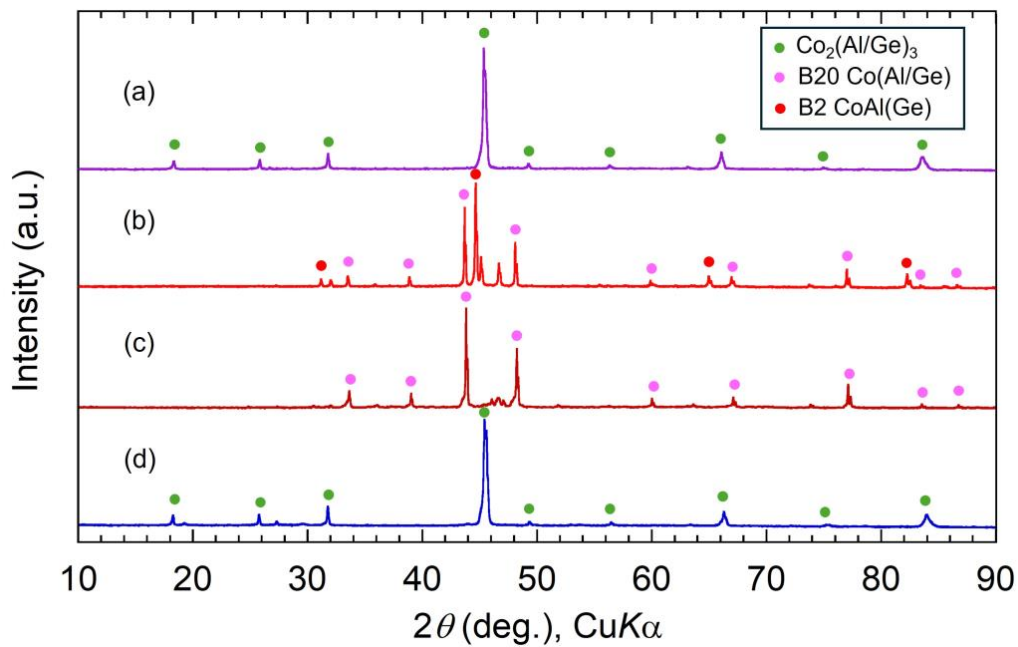

**Fig. S2** XRD patterns for the powder samples prepared with Co: Al: Ge = (a) 40: 30: 30, (b) 50: 20: 30, (c) 45: 10: 45, and (d) 35: 35: 30.

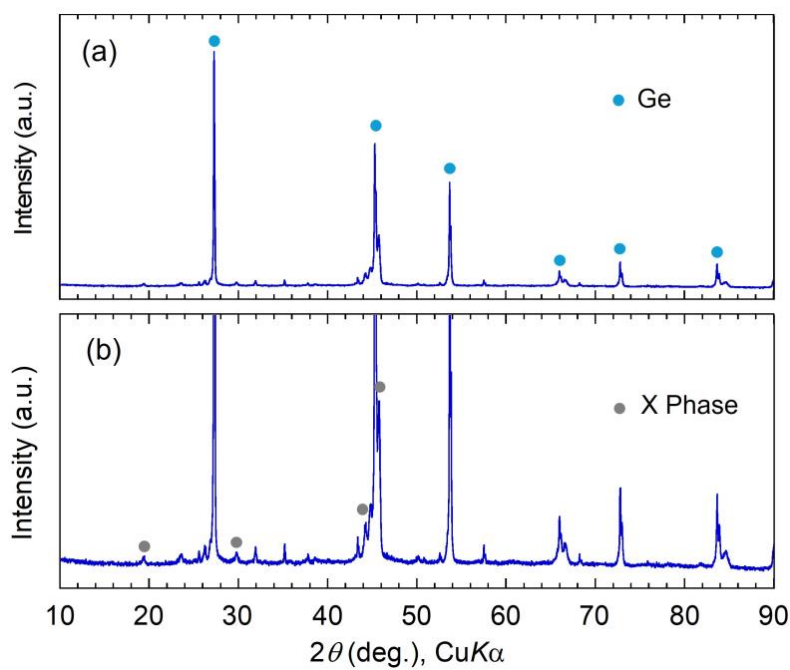

**Fig. S3** (a) XRD patterns for the powder samples prepared with Co: Al: Ge = 20: 40: 40. (b) An enlarged view of the upper XRD pattern with an adjusted upper limit.

**Supplementary Note B: Details of the unknown phase X**

To determine the crystal structure of the X phase, a single crystal suitable for X-ray diffraction measurement was prepared. The crystal was isolated from a sample synthesized with a molar ratio of Co: Al: Ge = 1: 2: 2. Single-crystal XRD intensity data were measured at room temperature using a single-crystal diffractometer (Bruker AXS, D8 QUEST, Mo-$K\alpha$). The software APEX3 [S1] was used to collect the diffraction data and refine the unit cell. X-ray absorption correction was performed using SADABS [S2] provided in APEX3. The SHELXL program [S3] installed with WinGX software [S4] was used to refine the atomic coordinates and displacement. The structures were drawn using the VESTA program [S5].

The XRD patterns from the single crystal were indexed with a cubic lattice parameter of $a$ = 11.2694(4) Å. The crystal structure was analyzed with the space group of $Im\bar{3}m$. The crystal structure of the X phase is shown in Fig. S4(a) and (b). The crystal structure consists of Co/Ge mixed-occupancy sites nested within a framework constructed from Ge/Al mixed-occupancy sites. The chemical formula of X phase determined by the XRD analysis was $Co_{29.68}Al_{70.02}Ge_{23.38}$. The result of the crystal structure analysis for the single crystals of the X phase is shown in Table S1.

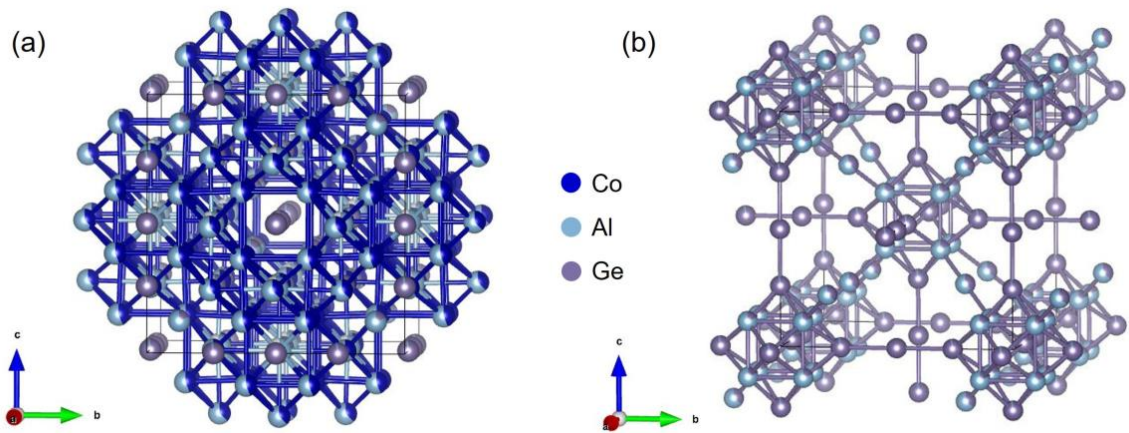

**Fig. S4** (a) Crystal structure and (b) the Ge/Al frame of the X phase.

**Table S1** Crystal data, data collection, and refinement of X-ray single crystal diffraction results for X phase.

| Chemical formula | $Co_{29.68}Al_{70.02}Ge_{23.38}$ |
|---|---|
| Formula weight, $M_r$ (g mol$^{-1}$) | 5335.34 |
| Temperature, $T$ (K) | 300(2) |
| Crystal system | Cubic |
| Space group | $Im\bar{3}m$ |
| Unit-cell dimensions, (Å) | $a$ = 11.2694(4) |
| Unit-cell volume, $V$ (Å$^3$) | 1431.21(15) |
| Z | 1 |
| Calculated density, $D_{cal}$ (Mg m$^{-3}$) | 6.190 |
| Radiation type | Mo $K\alpha$ |
| Crystal size (mm) | 0.041 ×0.043 ×0.070 |
| Absorption correction | Multi-scan |
| Absorption coefficient, $\mu$ (mm$^{-1}$) | 21.492 |
| Limiting indices | $-8 \leq h \leq 15$ |
| | $-14 \leq k \leq 13$ |
| | $-15 \leq l \leq 15$ |
| $\theta$ range for date collection, ° | 5.112–59.020 |
| No. of measured, independent and observed [$I > 2\sigma(I)$] reflections | 2444, 227, 183 |
| No. of reflections | 227 |
| No. of parameters | 27 |
| Weight parameters, $a$ / $b$ | 18.1972, 0 |
| $R1$, $wR2$($I > 2\sigma(I)$) | 0.0390, 0.0774 |
| $R1$, $wR2$(all data) | 0.0476, 0.0808 |
| goodness-of-fit on $F^2$, $S$ | 1.238 |
| largest diff.peak and hole, $\Delta\rho$ (eÅ$^{-3}$) | 0.505, −0.482 |

$R1 = \Sigma||F_o| - |F_c||/\Sigma|F_o|$. $wR2 = [\Sigma w(F_o^2 - F_c^2)^2/\Sigma(wF_o^2)^2]^{1/2}$, $w = 1/[\sigma^2(F_o^2) + (aP)^2 + bP]$, where $F_o$ is the observed structure factor, $F_c$ is the calculated structure factor, $\sigma$ is the standard deviation of $F_c^2$, and $P = (F_o^2 + 2F_c^2)/3$. $S = [\Sigma w(F_o^2 - F_c^2)^2/(n-p)]^{1/2}$, where $n$ is the number of reflections and $p$ is the total number of parameters refined.

**Supplementary Note C: LLM prompts and log files for experimental exploration**

In this note, we provide the prompts used for LLM-guided composition selection and the corresponding log files from each exploration cycle. The prompts inputted to Claude Code for generating composition proposals are summarized in Prompt S1, showing the cycle and strategy dependence of the prompt content. The complete log files for each cycle are provided as Log_S1.docx (Strategy A) and Log_S2.docx (Strategy B).

**Prompt S1:** Prompts for Claude Code depending on the strategies and cycles.

## Strategy A (aLLoyM + LLM) — Cycle 1

```
We want to construct a ternary phase diagram of the Co-Al-Ge system at 900°C
(1173.15K).

The file `candidates.csv` lists 231 candidate compositions on a 5% grid (colu
mns: Co, Al, Ge in fractions 0-1, phase column is empty).

The file `alloym_5pct_predictions.jsonl` contains phase predictions from aLLo
yM (a domain-specific LLM for alloy phase diagrams) for all 231 points. Each
line has Co_pct, Al_pct, Ge_pct (in %), and generated_answer (predicted phas
e).
However, these are merely predictions, and the phase names may not always be
accurate.

This is Cycle 1 — no experimental data exists yet. Using the aLLoyM predictio
ns as reference, select 8 compositions for the first round of experiments.

**Procedure:**
- Perform 10 independent selection runs. In each run, select 8 points with br
ief rationale.
- After all 10 runs, take a majority vote to determine the final 8 points.
- Document each run and the voting result.

Save the full selection process and rationale to `cycle1_method_a_log.md`.
```

## Strategy A (aLLoyM + LLM) — Cycle N (N ≥ 2)

We want to construct a ternary phase diagram of the Co-Al-Ge system at 900°C (1173.15K). This is cycle N.

The file `candidates.csv` contains 231 candidate compositions. Points that have been experimentally measured have their phase filled in the `phase` column. Points not yet measured have an empty `phase` column.

Then, using the experimental observations, select the next 8 compositions to measure.

**Procedure:**
- Perform 10 independent selection runs. In each run, select 8 points with brief rationale.
- After all 10 runs, take a majority vote to determine the final 8 points.
- Document each run and the voting result.

Save the full selection process and rationale to `cycleN_method_a_log.md`.

---

## Strategy B (LLM only) — Cycle 1

We want to construct a ternary phase diagram of the Co-Al-Ge system at 900°C (1173.15K).

The file `candidates.csv` lists 231 candidate compositions on a 5% grid (columns: Co, Al, Ge in fractions 0-1, phase column is empty).

This is Cycle 1 — no experimental data exists yet. Select 8 compositions for the first round of experiments.

**Procedure:**
- Perform 10 independent selection runs. In each run, select 8 points with brief rationale.
- After all 10 runs, take a majority vote to determine the final 8 points.
- Document each run and the voting result.

Save the full selection process and rationale to `cycle1_method_b_log.md`.

---

## Strategy B (LLM only) — Cycle N

We want to construct a ternary phase diagram of the Co-Al-Ge system at 900°C (1173.15K). This is cycle N.

The file `candidates.csv` contains 231 candidate compositions. Points that have been experimentally measured have their phase filled in the `phase` colum

n. Points not yet measured have an empty `phase` column.

Then, using the experimental observations, select the next 8 compositions to measure.

**Procedure:**
- Perform 10 independent selection runs. In each run, select 8 points with brief rationale.
- After all 10 runs, take a majority vote to determine the final 8 points.
- Document each run and the voting result.

Save the full selection process and rationale to `cycleN_method_b_log.md`.

---

**Supplementary Note D: LLM prompts for NIMO LLMRP module**

In this note, we provide the prompts used for LLM-guided composition selection in the benchmark test. These prompts are specified through the `prompt_file` and `system_prompt_file` parameters of the `nimo.selection` function, from which a single complete prompt for the Claude API is automatically constructed via NIMO package.

**Prompt S2:** Prompts for the Claude API used in the NIMO LLMEP module.

`system.md`

```
You are an expert materials scientist specializing in alloy phase diagram
construction.

Your task is to select the most informative experimental compositions from a
candidate list to efficiently map an unknown phase diagram.
```

`prompt.md` — Cycle 1

```
We want to construct a ternary phase diagram of the Co-Al-Ge system at 900°C
(1173.15K).

This is Cycle 1 – no experimental data exists yet. Select 8 compositions for
the first round of experiments.
```

`prompt.md` — Cycle N

```
We want to construct a ternary phase diagram of the Co-Al-Ge system at 900°C
(1173.15K).

Based on the experimental observations in the candidate table, select the
next 8 compositions.
```